\documentclass[conference]{IEEEtran}

\usepackage{soul}
\usepackage{amssymb,amsmath}
\usepackage{ifthen}
\usepackage{pifont}

\usepackage{soul} 
\usepackage{color}
\definecolor{KWColor}{rgb}{0.37,0.08,0.25}
\definecolor{CommentColor}{rgb}{0.133,0.545,0.133}
\definecolor{StringColor}{rgb}{0,0.126,0.941}

\newboolean{showcomments}
\setboolean{showcomments}{true}
\ifthenelse{\boolean{showcomments}}
{ 
  \newcommand{\mynote}[2]{
  \fbox{\bfseries\sffamily\scriptsize#1}
  {\small$\blacktriangleright$\textsf{\emph{#2}}$\blacktriangleleft$}}
}{ 
  \newcommand{\mynote}[2]{}
}

\usepackage{multirow}    
\usepackage{hhline}      
\usepackage{algorithm} 
\usepackage{algpseudocode}
\algnewcommand\algorithmicinput{\textbf{Input:}}
\algnewcommand\INPUT{\item[\algorithmicinput]}
\algnewcommand\algorithmicoutput{\textbf{Output:}}
\algnewcommand\OUTPUT{\item[\algorithmicinput]}

\usepackage{enumerate}
\usepackage{graphics}

\usepackage{cite}

\usepackage[font={small}]{caption}
\usepackage{subcaption}

\usepackage[hidelinks]{hyperref}
\usepackage[pdftex]{graphicx}
\usepackage{listings}
\usepackage{soul} 
\usepackage{color}

\definecolor{KWColor}{rgb}{0.37,0.08,0.25}
\definecolor{CommentColor}{rgb}{0.133,0.545,0.133}
\definecolor{StringColor}{rgb}{0,0.126,0.941}
\usepackage{balance}

\begin{document}

\author{
    \IEEEauthorblockN{Li Li, Tegawend\'e F. Bissyand\'e, Jacques Klein, Yves Le Traon}
    \IEEEauthorblockA{Interdisciplinary Centre for Security, Reliability and Trust, University of Luxembourg, Luxembourg\\
    \{li.li, tegawende.bissyande, jacques.klein, yves.letraon\}uni.lu}
}

\title{An Investigation into the Use of Common Libraries in Android Apps}

\maketitle

\begin{abstract}
The packaging model of Android apps requires  the entire code necessary for the execution of an app to be shipped into one single apk file. 
Thus, an analysis of Android apps often visits code which is not part of the functionality delivered by the app. 
Such code is often contributed by the common libraries which are used pervasively by all apps. 
Unfortunately, Android analyses, e.g., for piggybacking detection and malware detection, can produce inaccurate results if they do not take into account the case of library code, which constitute noise in app features. 

Despite some efforts on investigating Android libraries, the momentum of Android research has not yet produced a complete set of common libraries to further support in-depth analysis of Android apps.
In this paper, we leverage a dataset of about 1.5 million apps from Google Play to harvest potential common libraries, including advertisement libraries.
With several steps of refinements, we finally collect by far the largest set of  1,113 libraries  supporting common functionalities and 240 libraries for advertisement. We use the dataset to investigates several aspects of Android libraries, including their popularity and their proportion in Android app code. 
Based on these datasets, we have further performed several empirical investigations to confirm the motivations behind our work.
\end{abstract}

\section{Introduction}
\label{sec:introduction}

\begin{figure*}
    \centering
	\begin{subfigure}[b]{0.45\linewidth}
		\includegraphics[width=\linewidth]{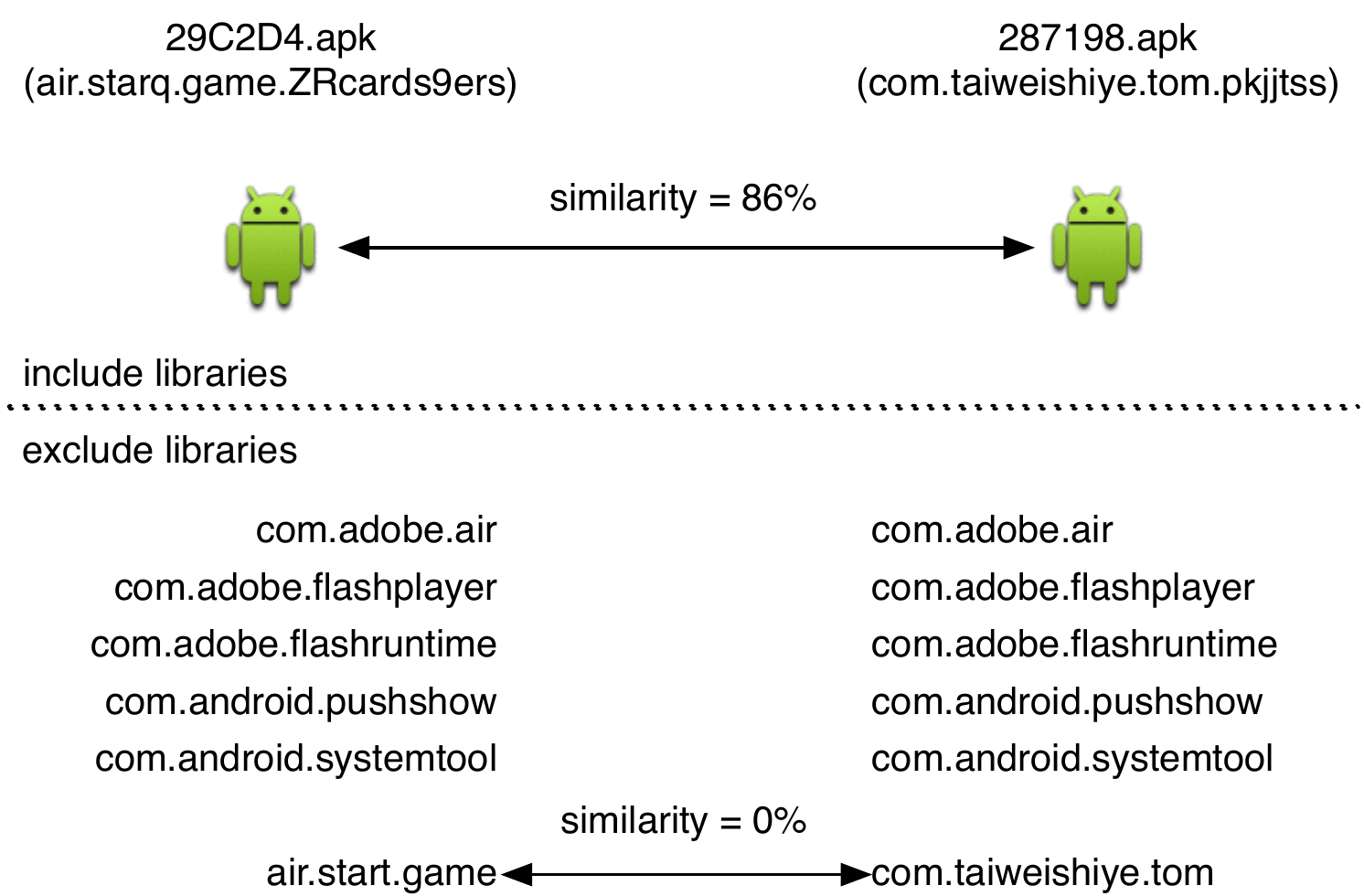}
		\caption{False Positive.}
		\label{fig:fp}
	\end{subfigure}\ \ \ \ %
	\begin{subfigure}[b]{0.45\linewidth}
		
		\includegraphics[width=\linewidth]{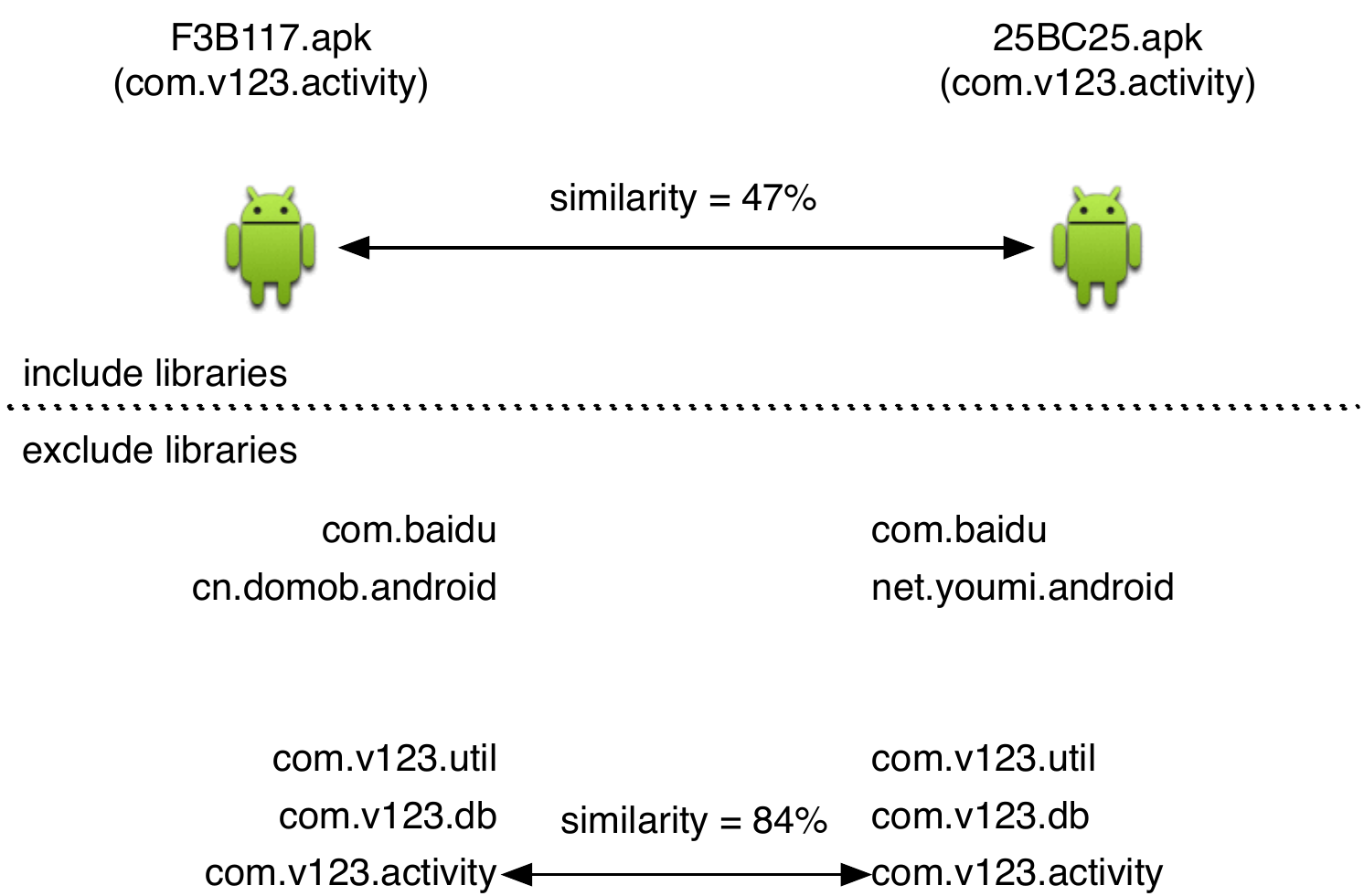}
		\caption{False Negative.}
		\label{fig:fn}
	\end{subfigure}
    
	\caption{Two motivating examples show the importance of excluding common libraries in order to perform precise piggybacked apps detection.
Note that \emph{F3B117.apk} and \emph{25BC25.apk} are actually signed by different certificates although they share a same package name.
	}   
    \label{fig:motivating_example}
\end{figure*}

Rapidly, Android has grown as a popular programming platform for developers and a worthwhile operating system
for manufacturers. 
In 2014 alone, over 1 billion of manufactured devices were equipped with Android, 
a significant gain from the 780.8 million shipments in 2013~\cite{android_shipments}. 
These devices, which, besides smartphones, include a number of household and office devices such as tablets, personal computers, TV sets, fridges, washing machines, etc., run a diversity of applications. 

Unfortunately, because these apps pervade all human activities, malicious or malfunctioning apps have become important threats that can lead to damages ranging from benign (e.g., app crashes) to critical (e.g., financial losses with malware sending premium-rate SMS, reputation issues with private data leaks, and potentially loss of human lives when apps will run on Android cars). 
These threats are further exacerbated in an ecosystem where thousands of applications written by hundreds of third-party developers are made readily available for download by users. 
Typically, as of July 2015, GooglePlay, the official market for free and paid Android apps, proposes over 1.6 million apps in various categories from productivity and messaging to games and social networking. 
Antivirus vendors, which regularly report on the status of malware spreading, have revealed that Android is now a target of choice for malicious attacks~\cite{SymantecReport2015}.
 
The research community has produced a large body of work for mitigating the emerged threats in the Android ecosystem, essentially to guard the security and privacy of users. 
For scalability and practicability reasons, a substantial number of the proposed approaches~\cite{li2015iccta,Amandroid,EnckTaintDroid} rely on static analysis 
to parse the entire code shipped in the app package to find security problems in code instructions, to extract features for further processing or simply to compare apps in large repositories. 
Unfortunately, because Android development paradigm allows to easily include third-party code, in the form of libraries, a significant portion of an app is eventually irrelevant to the functionality that it delivers. 
Common libraries embedded in app code thus constitute a significant barrier for the static exploration of applications code. 
There are indeed a number of research directions where tasks are hindered by the presence of common libraries in app code:

\noindent
{\em Repackaging detection.} Techniques for comparing apps to detect repackaged apps by computing their similarities
may provide inaccurate results when common libraries are pervasively used. 
In a preliminary study, Wang et al.~\cite{wang2015wukong} have found that over 60\% of Android apps' code is contributed by common libraries. 
To increase  accuracy in detection, most recent approaches have been considering filtering out such libraries, using heuristics.


\noindent
{\em Malware detection.} Recently, researchers have been focusing on machine learning techniques as scalable means to identify malicious apps in large datasets. 
To that end, they usually extract static features from the code. 
Unfortunately, the presence of library code may create significant noise making it hard to discriminate benign features from malware-specific ones. 
To account for such noise, some approaches, such as MUDFLOW~\cite{avdiienko2015mining}, assume that advertisement libraries, which are common libraries, are trustable. 
Thus, they simply ignore all results related to ad libraries, so as to focus on the real app code.

\noindent
{\em Code analysis.} Besides the false positives that may arise due to over-approximation, static code analysis is also often challenged by computing power and memory requirements. 
In the case of  FlowDroid~\cite{steven2014pldi}, the state-of-the-art static taint analysis tool for Android apps, it was reported that the analysis time can be too high~\cite{avdiienko2015mining}. 
Let us refer back again to Wang et al.'s findings, where 60\% of app's code are contributed by common libraries,
which would thus indicate that over half of the CPU and memory consumption is actually wasted on irrelevant library code, threatening the performance of the analyzer. 

The aforementioned cases constitute strong motivations for automatically identifying once a large set of 
\textbf{common libraries} from market-scale apps, which could then be used by other approaches to immediately take such libraries into account.
A straightforward solution for achieving such a task is to build a comprehensive \emph{whitelist} of common libraries.   
Wang et al.~\cite{wang2015wukong} claim to have collected more than 600 different common libraries to improve their repackaged app detection process. 
However, this collection is not available to the community, and may not be representative in other datasets. 
Other approaches~\cite{grace2012unsafe, pearce2012addroid, book2013case, gibler2013adrob} build on top of limited whitelists collected using simplistic heuristics and containing between only 9 (AdDroid~\cite{pearce2012addroid}) and 103 (Bootk et al.~\cite{book2013case}) libraries.

In this paper we investigate the use of common libraries in Android based on a dataset of around 1.5 million apps collected from the official Google Play market. In particular, we build and maintain a comprehensive whitelist of 1,113 Android common libraries that we share with the communities. Our approach identifies common libraries based on the assumption that they are used by many apps as such, i.e., without developer modification. 
We further label those libraries to distinguish between \textbf{advertisement libraries} (or ad libraries, a specific type of common libraries) and others, using heuristics defined from our manual investigations.


Overall, we make the following contributions:
\begin{itemize}

\item An approach to automatically harvest common libraries from market-scale Android apps.
In this work, we collect 1,113 common libraries from a dataset of around 1.5 million Android apps.

\item A discriminative study of advertisement libraries, for which 240 common libraries are recognized as ad libraries.

\item An empirical investigation and evaluation of the use of common libraries in Android apps.
We show that there are indeed significant differences in the use of common libraries between benign and malicious apps.
Besides, we also show that our harvested common libraries are indeed useful for other approaches, e.g., to reduce both false positive and false negative rates for piggybacked apps detection.

\item Two comprehensive \emph{whitelist}s of Android libraries (one for common libraries and the other for ad libraries), that we make available online to the Android research community at:\newline
\textsf{\small \url{https://github.com/serval-snt-uni-lu/CommonLibraries.git}}.

\end{itemize}

\section{Motivating Example}  \label{sec:example}
We now motivate our work by discussing the impact of filtering out libraries from apps when
performing piggybacking detection. 
Piggybacking is an operation that consists in taking an existing app, unpacking it, then modifying it by adding a (generally malicious) new payload and re-signing it, before distributing it as a new app. 
Like repackaged apps (where a payload is not necessarily added), piggybacked apps are now pervasive in the Android ecosystem where they further constitute an easy way to build and  distribute malware~\cite{zhou2012detecting,zhou2013fast,allix2014forensic}. 
A typical approach for detecting piggybacked apps consists in performing pairwise comparisons to identify the original app that was actually piggybacked. In the process of computing similarity however, libraries, which may account for a large portion of apps, can influence towards inaccurate results. We present two real-world examples of pairs of apps where the presence of libraries can lead to a mislabelling of a legitimate app as piggybacked or a failure to flag a piggybacked app as such. 


\subsection{Mislabeling Legitimate apps as Piggybacked}
We consider in Fig.~\ref{fig:fp} the case of two apps (\emph{air.starq.game.ZRcards9ers} and \emph{com.taiweishiye.tom.pkjjtss}) collected from an Android market. 
The packages in their code structure are very similar when considering the common libraries that they integrate: one app has 86\% of its code\footnote{The percentage is computed based on method level, where more details will be given in Section~\ref{subsec:step2}.} that is also contained in the other app.
However, considering the results of a prior investigation  of a set of 1,169 known legitimate/piggybacked app pairs where we found that most of the similarity degree ranges between 81\% and 100\%, 
we could set a threshold of 80\% for identifying piggybacking cases. This unfortunately would lead to a mislabeling in the above case. 
Indeed, a detailed analysis of both apps shows that they are actually using several common libraries (e.g., {\tt com.android} and {\tt com.adobe}). Excluding such libraries from the similarity computation, the similarity degree falls down to 0\%, leaving no room for a false positive prediction.


\subsection{Missing True Piggybacked Apps}
We now consider in Fig.~\ref{fig:fn} two apps which are known to be a legitimate/piggybacked app pair. 
These apps share the main package called \emph{com.v123.activity}. 
However, library \emph{cn.domob.android} was replaced in the piggybacked app with library \emph{net.youmi.android} to redirect the revenues of the legitimate app to another developer. 
Nevertheless, although these two apps are piggybacked from one to another, their similarity degree is only at 47\%, which would constitute a false negative in our detection scheme with a  threshold at 80\%. 
However, if the detection system identified first the common libraries and dismissed them during pairwise comparison, the similarity degree would reach 84\%, leading to a successful prediction.

Overall, the validity of pairwise comparison for piggybacking detection could be threatened when substantial parts of app code are common library code. 
Thus, to limit both false positives and false negatives, library filtering is now more and more considered in state-of-the-art repackaging and piggybacking detection approaches~\cite{wang2015wukong, hanna2013juxtapp, chen2014achieving}. 
However, the whitelists that they leveraged is built based on manual investigations or automatically with limited datasets. 
Furthermore, these whitelists are seldom available to other researchers in the community.

\vspace{0.2cm}
\begin{tabular}{|p{3.1in}|}
\hline {\em \textbf{One objective} of our work is to provide to the community a comprehensive list of common libraries, which can be used as a whitelist for supporting static code-based analyses of Android apps.
}\\ \hline
\end{tabular}
\vspace{0.5cm}

\vspace{-3mm}
\section{Identification of Common Libraries} \label{sec:approach}
In this section we provide details on the approach that we have devised to collect common libraries for the study. 



\begin{figure}[!h]
\centering
\includegraphics[width=\linewidth]{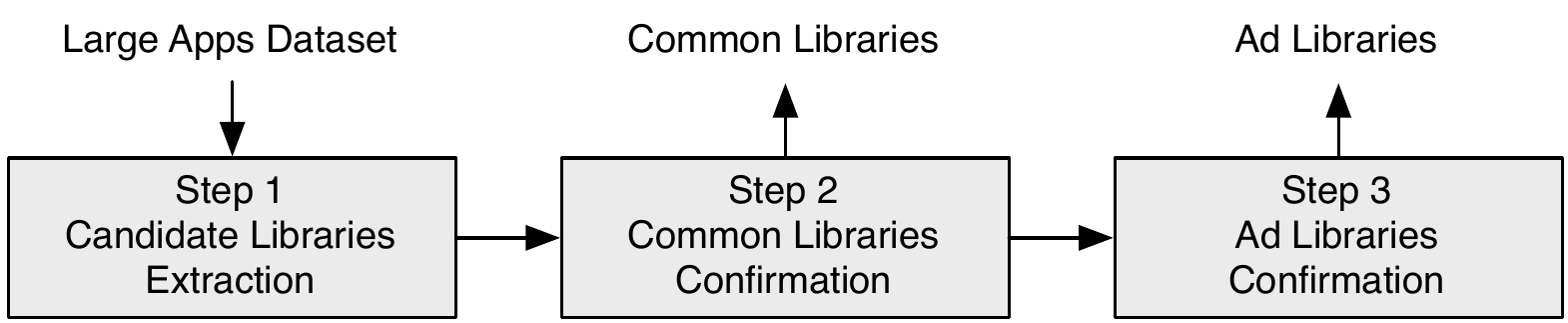}
\caption{Approach overview.}
\label{fig:overview}
\end{figure}

\textbf{Process Overview:} Fig.~\ref{fig:overview} illustrates the general process of our approach,
which is dedicated to harvest common libraries in Android apps and identifying advertisement libraries among them. 

\vspace{-3mm}
\vspace{-3mm}

First, for our approach to make sense, we need a large and representative dataset of Android apps. 
Then, as a first step, we visit all the apps in the dataset and rank all packages in terms of the frequency of their appearance in apps. For the sake of simplicity, we assume that a package with the same name in several apps is a candidate library. Thus, {\em Step 1} outputs a ranked list of candidate libraries, where the highest ranked candidate library has the most recurring package name in the dataset.
In the second step we perform a more fine-grained pairwise comparison of candidate library code within apps. The objective of {\em Step 2} is to confirm as common library packages those recurring packages that have the same name and are very similar in their code.
Finally, in  {\em Step 3}, we further investigate the harvested libraries to label those that are advertisement libraries and thus may be treated differently in some Android analysis approaches.
 
We now provide details on how each step works in the following three subsections.

\subsection{Step 1: Candidate Libraries Extraction}
\label{subsec:step1}
We assume that common libraries are such software packages that are:
\begin{itemize}
	\item used in a large number of apps -- recurring packages have very high probability of being common libraries.
	\item used by developers without modifications -- their code must be similar across apps. Hu et al.~\cite{hu2014duet} have found that over 80\% of libraries are indeed used without modification in their dataset of 100,000 Google Play apps. 
\end{itemize} 

Building on those assumptions, and leveraging a large dataset, we extract all package names from Android apps and cluster them based on their frequency of occurrence in the dataset. Theoretically, packages that appear in at least two apps could be taken as candidate libraries\footnote{Actually this may not be true if the apps are from a same developer. However, since we are performing experiments on a large set of apps, this small deviation will not impact our final results.}. To reduce the number of distinct packages considered as candidate libraries, and which must be further processed we consider two constraints:
\begin{itemize}
	\item We only consider the first three segments\footnote{In this paper, we use the term segment to describe each domain of different levels, e.g., for package \emph{org.example}, we say it contains two segments, which are \emph{org} and \emph{example}.} of package name or the entire name if there are less than 3 segments.
 With this constraint we manage to limit the number of redundant subpackages while still guaranteeing a large diversity in package names. 
	\item We also exclude packages with names starting with {\em  android.support}. Indeed, there are many sub-packages within this package and they are used pervasively in Android apps. Furthermore, since these are part of the Android framework, we do not consider them in our study.
\end{itemize}


\subsection{Step 2: Common Libraries Confirmation}
\label{subsec:step2}

\begin{figure*}
\centering
\includegraphics[width=\textwidth]{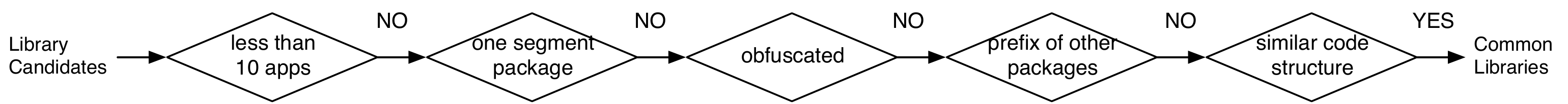}
\caption{Refinement process for common libraries identification}
\label{fig:conformation}
\end{figure*}
Because package naming is done in Java programming with limited constraints, any two packages may share the same name while being completely different in terms of code functionalities. Also, the frequency of a package name may actually be contributed by piggybacking operations, obfuscation activities (e.g., \emph{a.a.a} or \emph{com.a} are recurrent in many obfuscated apps) or simplistic naming (e.g., \emph{debug} or \emph{mobile} package names). Thus, we must refine the list collected in the previous step with code similarity measurements to find actual code packages used as common libraries.

Beforehand, given the expensive property of pairwise comparison, we use heuristics to exclude from the candidate libraries outputted by {\em Step 1}, those packages which would be irrelevant.
Our refinement process is shown in Fig.~\ref{fig:conformation}.

1) At first, we focus on those packages whose names appear in more than 10 apps to reduce the number of candidate libraries to the most relevant ones. 

2) Then, we remove such packages whose names contain only one segment. Although such short names are indeed likely to be redundant in several apps, they are not likely to be those of packages that will be distributed as common libraries. Indeed, to prevent package name collisions, one convention in Java package naming\footnote{\url{https://newcircle.com/bookshelf/java_fundamentals_tutorial/packaging}}  recommends organisations/development teams to use their reversed Internet domain names (e.g., com.facebook) to begin their package names, which justifies our assumption that common libraties, intended for wide distribution, have package names with several segments. 

3) Next, we undertake to exclude packages with obfuscated code. However, because there is currently no advanced approach for checking whether a package is obfuscated or not, we build on a naive approach  based on observations of several obfuscated apps: every package that contains a single letter segment (e.g., $d$ of \emph{com.idreamsky.d}) is considered as obfuscated. 
%

4) To further reduce the number of candidates, we exclude such packages that are prefixes of other packages (e.g., we remove package \emph{com.sansec} if package \emph{com.sansec.AESlib} exists).
The idea behind this decision is that on the one hand long packages would indicate more fine-grained examination while, on the other hand, short packages would increase the chance of being duplicated (by accident).

5) Finally in the last step, we perform package similarity analysis to discriminate common library packages from normal app code package. Given $p$, a package name, and $A$, a set of apps which include a package named $p$, our similarity analysis works in three steps:


\begin{itemize}
 \item {\em Pairwise combinations of apps.} We consider all the pairwise combinations of apps with package name $p$. Recall that every considered package name $p$ was selected as candidate because it appears in at least 10 apps. Thus, for any given $p$, there are at least  ${10 \choose 2} = 45$ pairs to compare. Google's ad package {\em com.google.ads} is the one for which $A$ is the largest (247,394 apps), leading to a over 30 billions pairs that require comparisons. For scalability reasons, we randomly select for each case of a package name $p$, 10 pairs of apps, allowing us to assess whether this package name indeed represents a common package code across the apps.


\item {\em Method Comparisons.} Analysis of a pair of apps is performed by computing the similarity between their methods. This similarity takes into account not only the signatures of apps but also their respective contents. Two methods, from two different apps, with the same signature are said to be {\em identical} only when their contents are the same. Otherwise, they are simply said to be {\em similar}. Such methods may exist between two packages of the same library in several cases: a method in one library package may been modified to insert malicious payload during piggybacking operations; different obfuscation algorithms applied on different apps that include the same library may produce methods with the same signature but different contents. To limit the impact of obfuscation, we proceed to abstract the contents of methods by comparing the types of statements (e.g., ``invoke") in the Jimple code, leaving out all names of variables/fields/methods. However, since obfuscation is not expected to modify SDK API methods, we also take into account the names of such methods. Eventually, the similarity of methods is computed as a simple text differencing.
\item {\em Similarity Analysis.}

In the last step, we finally perform pairwise similarity analysis for
packages with the same name $p$. 
There are two thresholds, namely  $t_p$ and $t_a$, which are involved in
the similarity analysis. First, we consider that two packages $p_1$ and
$p_2$ correspond to the same common library $p$ if $p_1$ and $p_2$ are
identical or are at least similar up to a threshold $t_p$.
Second, because of the known common phenomenon of repackaging/piggybacking
in Android, which may nullify the package similarity (because they are probably from the same original app), we must dismiss cases where a the similarity score of the pair of apps ($app_1$ and $app_2$) is higher than a threshold~$t_a$.

Note that the similarity between apps is computed at the method level (i.e., what percentage of methods are identical or similar between the apps?).
\end{itemize}

To summarize, for similarity analysis, given a pair of apps ($app_1$, $app_2$), we compute  four metrics: \emph{identical} (i.e., the number of methods that are exactly the same, both in terms of signatures and implementation), \emph{similar} (i.e., the number of methods having the same signature but with different contents),  $deleted$ (i.e., the number of methods that exist in $app_1$ but not in $app_2$), and \emph{new} (i.e., the number of methods existing only in $app_2$). 
These metrics are indeed good indicators for comparison and have been leveraged in state-of-the-art Android similarity tools, such as Androguard~\cite{desnos2012android}. 
Given these metrics, we can compute the similarity between the pair ($app_1$, $app_2$) using Formula~\ref{eq:similarity}. 
\vspace{-1mm}
\begin{equation}\label{eq:similarity}
\small
similarity = max\{\frac{identical}{total - new}, \frac{identical}{total - deleted}\}
\end{equation}

where 
\vspace{-3mm}
\begin{equation}
\small
total = identical + similar + deleted + new
\end{equation}
\vspace{-1mm}
Note that we use the same formula to perform the similarity analysis of a given pair of packages $(p_1, p2)$, except that the metrics  are computed by counting methods in packages rather than in apps (e.g. $identical$ is the number of methods that are exactly the same in $p_1$ and $p2$, $deleted$ is the number of methods that exist in $p_1$ but not in $p_2$, etc.)

%

\vspace{-1mm}
\subsection{Step 3: Identification of Ad Libraries}
\label{subsec:step3}
A specific example of type of widespread common libraries in Android is advertisement libraries.
Such libraries are indeed used pervasively as they constitute one of the main ways for app developers to be rewarded for their development effort. 
Ad libraries are also often inserted during piggybacking to redirect revenues. 
Their presence in an app also often lead antivirus products to flag them as adware. Recent approaches for Android security analysis are now processing ad library code in a specific way to reduce false positives. 
For example, MUDFLOW~\cite{avdiienko2015mining} simply does not report any potential sensitive data leaks through ad libraries, as they might be legitimate. To that end, they have leveraged a limited whitelist of 12 libraries. 
In this context, we propose to further mine our collected set of common libraries to identify a large set of ad libraries which could be leveraged to improve the results of Android analyses. 
To that end, we consider a basic method of detection based on the library name and a more semantic approach based on the characteristics of ad libraries.

\subsubsection{Keywords matching} We note that ad library package names generally contain keywords that include the term {\em ``ad''}. Widespread examples of such packages are {\em com.google.ads} and {\em com.adsdk.sdk}.
Unfortunately, simply matching ``ad" in the package name would lead to substantial portion of false positives as several library package names have ``ad" in their segments which are common words (e.g., shadow, gadget, load, adapter, adobe). 
Thus, to work around this limitation, we collect all English words containing ``ad" from SCOWL\footnote{Spell Checker Oriented Word Lists: http://wordlist.aspell.net} (accounting for a total of 13,385 words), and dismiss packages containing such words as potential ad libraries.

\subsubsection{Ad features investigations}

We consider samples from a list of ad packages summarized by Grace et al.~\cite{grace2012unsafe} and manually investigate how ad libraries differentiate from other common libraries, and infer a set of features whose presence in a package would justify the tag of ad library.

\paragraph*{Internet usage}
All investigated libraries unsurprisingly require access to Internet to remotely upload to a server some viewing statistics and update ad contents. Thus, apps integrating add libraries also require the \texttt{android.permission.INTERNET} permission. Given this fact, we can already exclude a number of common libraries, which appear in apps without Internet access, as ad libraries. 
However, given that an app may requests the \emph{INTERNET} permission for its own needs, we cannot immediately state that a common library in such an app is an add library. Instead we must investigate whether the code of such an app indeed declares uses Internet-related APIs. To that end, we leveraged the whitelist of such APIs, originally shared by PSCout~\cite{pscout}, to produce candidate ad libraries among the common libraries.

\paragraph*{Components declaration}
Our manual investigations have also revealed that ad libraries often contain components, mainly Activities, for facilitating users' ad-related interactions (e.g., switching to a new full-screen ad page  when users click on an advertisement banner). As a concrete example, MoPub\footnote{https://github.com/mopub/mopub-android-sdk} is an advertisement library targeting  both Android and iOS. To integrate this library in their apps, developers must declare four components in their apps' \emph{manifest} file. One component in particular, \emph{MraidVideoPlayerActivity} is necessary for video ads to work properly. Thus, when a library package is associated to a declared component, we flag it as a potential ad library.

\paragraph*{Views declaration}
In Android, advertisements are generally set to be visualized, which from in Android programming imply the use of view gadgets (i.e., classes extended from \texttt{android.view.View}).
Thus, we check whether there are \emph{View}-based classes under a common library to flag it as candidate ad library.

\subsection{Implementation details}    \label{subsec:implementation}
We implement our approach through several languages such as Java and shell/python scripts.
In \emph{step 1}, we leverage \emph{Apktool}\footnote{https://ibotpeaches.github.io/Apktool/} to disassemble Android apps.
Given an android app, we extract the prefixes of paths of \emph{smali} files (a format used by \emph{Apktool} to represent Android apps' code) to represent its packages.
Then, we cluster all the packages of investigated apps together and rank them through their repeated times.
The packages whose size are greater than a given threshold are selected as library candidates.

The code similarity analysis in \emph{step 2} and the ad library conformation in \emph{step 3} are implemented in Java.
More specifically, both of them leverage Soot~\cite{lam2011soot} to achieve their functionality and work in the \emph{Jimple} code level, where Soot is a framework for analyzing and transforming Java/Android apps while Jimple is an intermediate representation of Soot.
The transformation from Android Dalvik bytecode into Jimple code is powered by Dexpler~\cite{bartel:soap2012}, which currently is available as a plugin in Soot.

\section{Dataset and Results}
\label{sec:results}

In this section, we first disclose our evaluated data set in Section~\ref{subsec:dataset} and then we present our overall findings including both common libraries and also ad libraries in Section~\ref{subsec:overall_results}. Finally, we present further statistics on the libraries in Section~\ref{subsec:popularity} and Section~\ref{subsec:portion}.

\subsection{Dataset}		\label{subsec:dataset}


Our data set is made up of 1,455,516 (around 1.5 million) apps that are collected from the official Google market (\textit{Google Play}) over several months. This data set has already been applied for large-scale experiments on Android researches such as malware detection~\cite{allix2014forensic, allix2014empirical, li2015potential} and piggybacked apps detection~\cite{li2015ungrafting}.
We have sent all the apps into VirusTotal to check whether they are malicious or not.
Among the 1,455,516 apps, 311,490 (nearly 21\%) of them are flagged by at least one anti-virus product hosted on VirusTotal while 65,079 (nearly 4\%) apps are flagged by at least five anti-virus products.


\subsection{Overall Results} 	\label{subsec:overall_results}

\begin{table}[!h]
    \caption{Summary of our investigation results.}
    \label{tab:results1}
    \centering

    \begin{tabular}{  l  r }
        \hline
Type											& Number  \\ \hline
\#. of packages (total)						& 7,710,505\\
\#. of packages (distinct)					& 676,674 \\
\#. of packages ($N_{shared\_apps} > 10$)	& 19,725 \\
\hline
\#. of packages (one segment)				& 613 \\
\#. of packages (obfuscated)					& 1,461 \\
\#. of packages (prefix of others)			& 919 \\
\hline
Size of final set of candidate common libraries			& {\bf 16,732} \\
        \hline
    \end{tabular}
\end{table}

Table~\ref{tab:results1} illustrates the overall results of our investigation on a data set of around 1.5 million apps.
In total, we collect 676,674 distinct package names, where we filter out 656,949 package names that are used by at most 10 apps, leading to a set of 19,715 package names.
We further dismiss 2,993 from consideration thanks to our library refinement process.
Those 2,993 package names are composed of 613 one segment packages, 1,461 obfuscated packages and 919 packages that are prefix of other packages.
Finally, we perform pairwise similarity analysis for 16,732 packages.
For each package, we randomly select 10 pairs of apps to do the comparison.
As long as there are positive results, we consider it as a common library, and verse visa.


\subsubsection{Results of Common Libraries}

\begin{table}[!h]
    \caption{Results of common libraries with different thresholds: $t_p$ for package-level and $t_a$ for app-level. Common libraries are select if and only if their package-level similarities are bigger than $t_p$ while their app-level similarities are smaller than $t_a$.}
    \label{tab:results2}
    \centering

    \begin{tabular}{  l  c c c c }
        \hline
$t_p$\textbackslash $t_a$		& 0.1		& 0.2	& 0.3	& 0.4 \\ \hline
0.9								& 1,113		& 2,148		& 3,173		& 4,072	\\
0.8								& 1,363 		& 2,564		& 3,715		& 4,685	\\
0.7								& 1,573		& 2,898		& 4,117		& 5,144	\\
0.6								& 1,735		& 3,179		& 4,452		& 5,509	\\
        \hline
    \end{tabular}
\end{table}

Our common libraries selection is actually depending on the two thresholds introduced in Section~\ref{sec:approach}: $t_a$ for app-level similarity and $t_p$ for package-level similarity.
The precision of our results is positively correlated to $t_p$ while negatively correlated to $t_a$. 
Indeed, 
the bigger $t_p$ is, the higher the probability that a given candidate library is an actual common library, 
giving the assumption that libraries are not modified when they are used among apps.
 On the other hand, the smaller $t_a$ is, the lower the probability that the compared two apps are repackaged/piggybacked from one to another. 
Recall that if two apps are repackaged/piggybacked from one to another, the similarity of packages would become meaningless, 
as in this case, most packages would be the same, without being necessarily common libraries.

Table~\ref{tab:results2} illustrates the results of common libraries with different thresholds.
The final number of common libraries range from 1,113 to 5,509.
To better refer to our results in the remainder of the paper, we name $CL_{p,a}$ the set of Common Libraries that are selected with the thresholds $t_p$ and $t_a$.
For example, $CL_{9,1}$ stands for the precise set of 1,113 common libraries we harvest with $t_p=0.9$ and $t_a=0.1$, while $CL_{6,4}$ stands for the more ``loose" 
set of common libraries, 
which although is the biggest set, contains potential more false positives (less precise than $CL_{9,1}$).

\subsubsection{Results of Ad Libraries}

\begin{table}[!h]
    \caption{Results of ad libraries.}
    \label{tab:adlibs}
    \centering

    \begin{tabular}{ l r }
\hline
Description 	&	\#. of Libraries	 \\
\hline
Ad-related keyword matching	&	275 \\
Ad characteristic-based investigating	& 822 \\
Merge (conservative ad libraries)		& 1050 \\	
\hline
\hline
Manual confirmation (keyword matching)	&	222	\\
Manual confirmation	(characteristic investigating)	& 137	\\
Merge (precise ad libraries)				& 240 \\
\hline
    \end{tabular}
\end{table}

We then distill ad libraries from the previously harvested common libraries.
We start from the $CL_{6,4}$ library set and performs two types of refinement: 
1) ad-related keywords matching and 2) ad characteristic-based investigation.
The refinement results are presented in Table~\ref{tab:adlibs}.

\textbf{Ad-related keywords matching.}
By following the process described in Section~\ref{subsec:step3}, 
we were able to automatically harvest 275 ad libraries.

\begin{figure}[!h]
\centering
\includegraphics[width=0.8\linewidth]{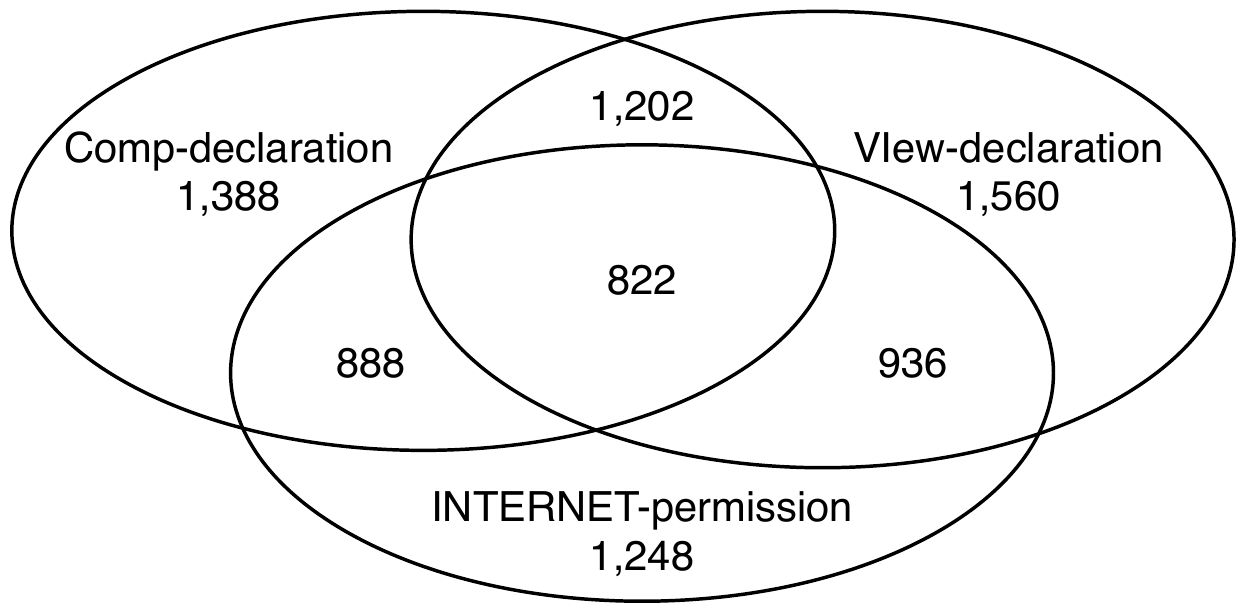}
\caption{Investigation results of different characteristics for ad libraries.}
\label{fig:characteristic}
\end{figure}

\textbf{Ad characteristic-based investigating.}
We have observed three characteristics that ad libraries may have in Section~\ref{subsec:step3}.
Fig.~\ref{fig:characteristic} shows the results of our investigation.
Among the 5,509 libraries in $CL_{6,4}$, 1,248 of them request the \emph{INTERNET} permission, 1,560 have declared \emph{View} gadgets and 1,388 have declared components.
The intersection results are also illustrated in Fig.~\ref{fig:characteristic}.
In this work, we take the intersection of all the three characteristics as potential ad libraries, leading to a set of 822 ad libraries.

In the next step, we merge the aforementioned two ad libraries sets, leading to a sef of 1,050 ad libraries.
In the remainder of the paper, we name this set $AD_{1050}$.

\textbf{Manual confirmation.}
As far as we know, $AD_{1050}$ is currently the largest set of ad libraries existing in the community.
However, because we start from $CL_{6,4}$, mainly to start with the biggest set (minimizing the miss of libraries), $AD_{1050}$ may contain false positives.
To this end, we perform a fast but aggressive manual refinement, where only clear ad libraries\footnote{Their corresponding web pages have explicitly claimed that they function advertisements.} are taken into account.
As a result, 240 libraries are confirmed as ad libraries\footnote{This does not mean the remaining 810 libraries are not ad libraries.}, hereinafter we refer to this set as $AD_{240}$.
This 240 ad libraries are highly precise.
We argue that a highly precise ad library set is important, which plays as a basement that makes it possible for other approaches to also yield precise results.

\subsection{Popularity of common libraries}
\label{subsec:popularity}

\begin{figure}[!h]
\centering
\includegraphics[width=\linewidth]{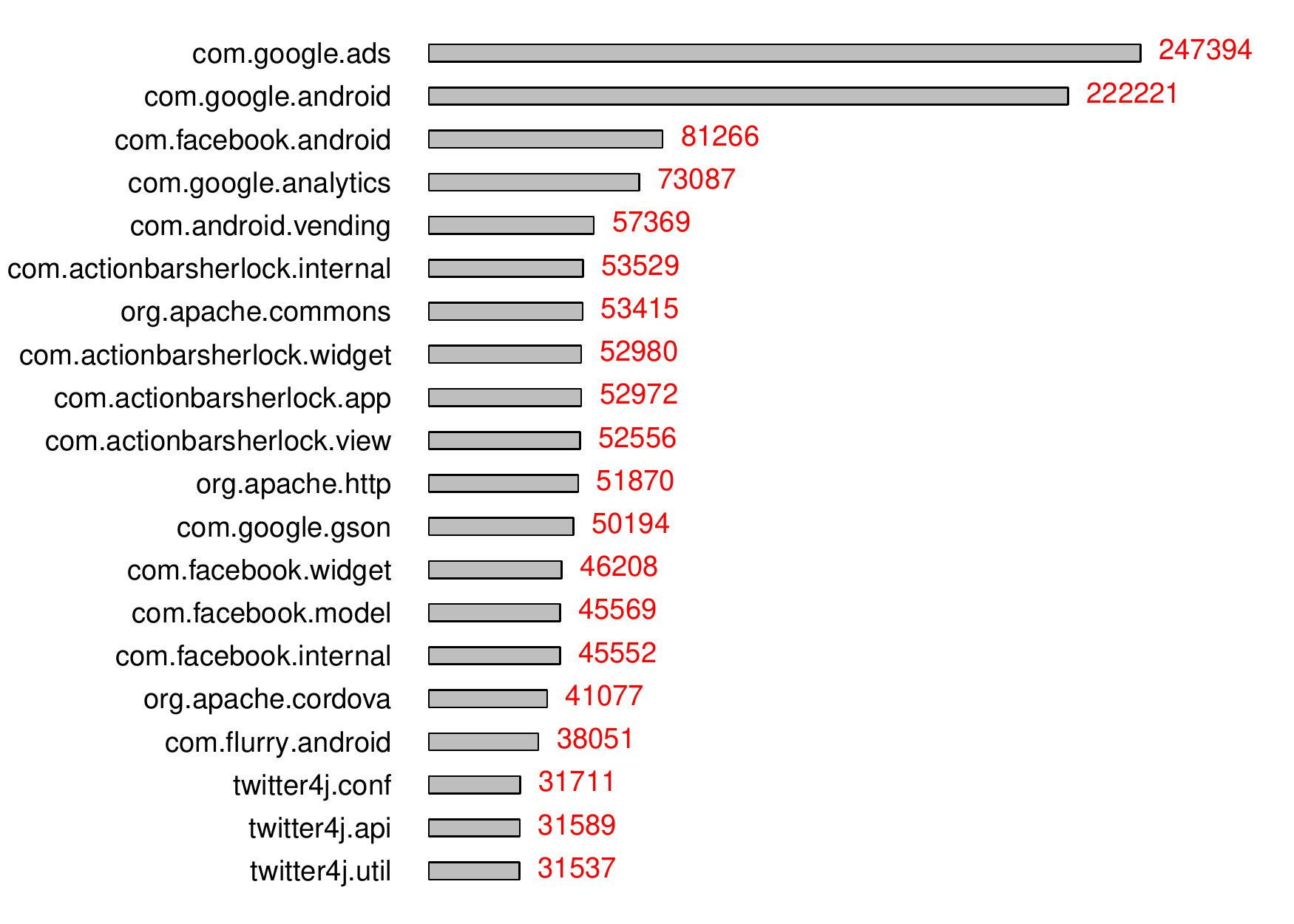}
\caption{Popularity of the top 20 common libraries in our investigation, and the number of apps in which they are used.}
\label{fig:popularity}
\end{figure}

Fig.~\ref{fig:popularity} lists the top 20 common libraries and indicates, for each, the number of apps in which they are used.
The top used library is \emph{com.google.ads}, which is used by 247,394 apps (nearly 17\%) of our data set. 
Moreover, the results suggest that developers often use libraries which are proposed by popular (well-known) companies such as Google or Facebook.

\vspace{0.2cm}
\begin{tabular}{|p{3.1in}|}
\hline {\em 
The most used common library in Android apps is {\em com.google.ads}, an adverstisement library included in nearly 17\% of apps. 
}\\ \hline
\end{tabular}


\subsection{Proportion of Library code in App code}
\label{subsec:portion}
We then look into the percentage of Android apps code which come from libraries. 
To this end, we consider the libraries present in $CL_{9,1}$ and a  set of 10,000 apps randomly selected from our initial set of apps. 
For each app, we compute the size of the $CL_{9,1}$ libraries ($size_{lib}$, in bytes) presented in the app and the size of the whole app ($size_{app}$). 
We finally compute the portion $p$ of the use of common libraries through $p = size_{lib}/size_{app}$. 
The experimental results vary from 0 to 0.99, giving a median value 0.41.
Among the 10,000 apps, 4,293 (42.9\%) of them have used more code in libraries than in their real logic ($p >= 0.5$).
This results show that Android apps are indeed using common libraries pervasively.

\vspace{0.2cm}
\begin{tabular}{|p{3.1in}|}
\hline {\em 
42\% of our sampled app packages contain more common library code than specific app code. On average, 41\% of an Android app code is contributed by common libraries.
}\\ \hline
\end{tabular}

\section{Empirical Investigations}
\label{sec:evaluation}
Beyond the initial goal to provide a comprehensive and publicly accessible \emph{whitelist} of Android libraries, we investigate some potential benefits of having such a list. In particular, based on the collected dataset of libraries we study the following research questions:
\begin{itemize}
\item Are libraries used similarly by benign and malicious apps? Can they be used as naive features for learning anti-virus predictions?
\item What improvements to Android analysis can be brought with an access to our harvested libraries? In particular what is the impact of taking into account our set of libraries on the performance of piggybacking detection and machine learning-based malware detection?
\item To what extent is the set of collected ad libraries exhaustive compared to baseline from VirusTotal?

\end{itemize}


\begin{figure}
    \centering
	\begin{subfigure}[b]{0.49\linewidth}
		\includegraphics[width=\linewidth]{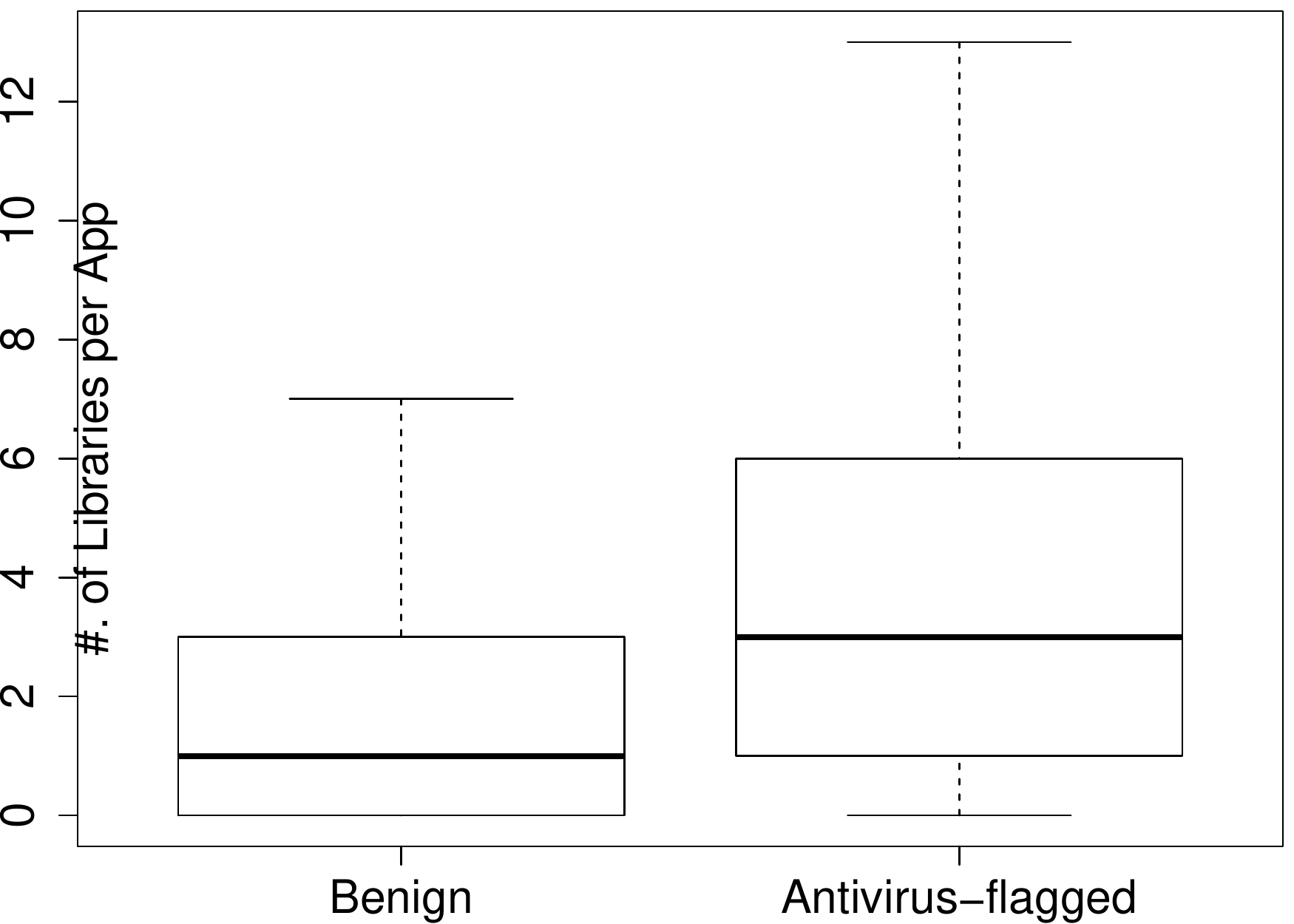}
		\caption{Common libraries.}
		\label{fig:boxplot_rq1}
	\end{subfigure}%
	\begin{subfigure}[b]{0.49\linewidth}
		
		\includegraphics[width=\linewidth]{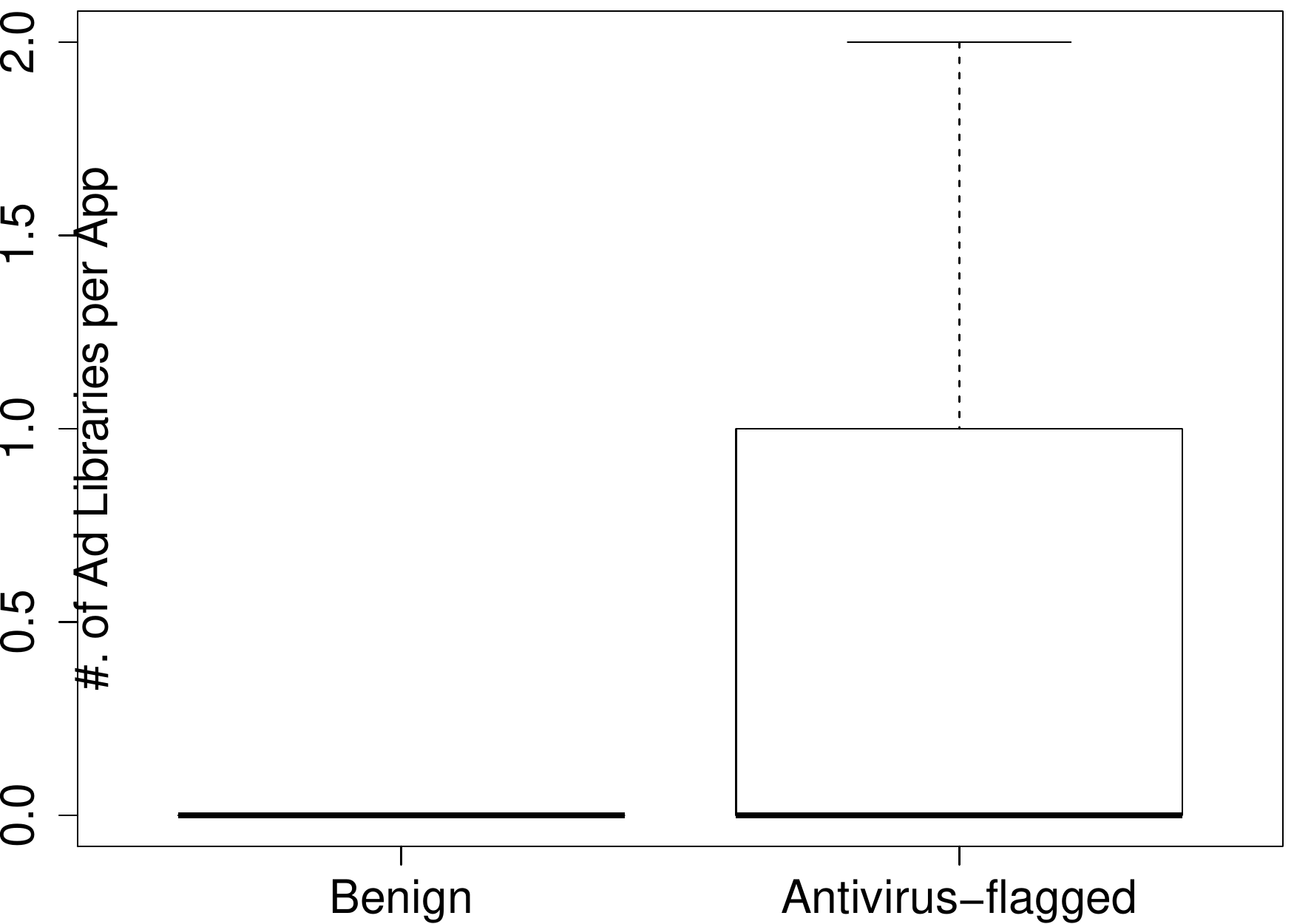}
		\caption{Ad libraries.}
		\label{fig:boxplot_rq1_ad}
	\end{subfigure}
    
	\caption{Library usage between benign and antivirus-flagged apps}   
    \label{fig:motivating_example}
\end{figure}

\subsection{RQ1: Benign vs. Antivirus-flagged Apps} \label{subsec:rq1}
With this research question, we are interested in investigating the differences in library usage 
between benign and antivirus-flagged apps. To this end, we randomly select 20,000 apps, 10,000 benign  and 10,000 flagged by at least one anti-virus product hosted by VirusTotal\footnote{https://www.virustotal.com}.

Among the 20,000 apps, 5,424 benign ones and 8,580 flagged ones use at least one common library.
In total, 892 out of the 1,113 $CL_{91}$ common libraries are used.
Figure~\ref{fig:boxplot_rq1} shows the boxplot with the number of common libraries used by each app in both categories. The median value for benign apps is 1 whereas the median value is 3 for antivirus-flagged apps. We confirm that this difference is significant by performing a Mann-Whitney-Wilcoxon (MWW) test. Benign apps thus use less libraries than anti-virus flagged ones. 


We further study whether this is related to advertisement libraries. Among the 240 libraries in $AD_{240}$, 98 of them are used by 1,332 benign apps and 3,209 antivirus-flagged apps. Figure~\ref{fig:boxplot_rq1_ad} shows the boxplot which suggests that antivirus-flagged apps contain significantly more ad libraries than benign apps.


In addition to the quantitative comparison, we investigate whether the appearance of the collected common libraries can be used to discriminate antivirus-flagged apps\footnote{In this work, we consider those antivirus-flagged apps as potentially malicious.} from benign apps through machine learning-based malware classification.
To this end, we leverage \emph{RandomForest} algorithm~\cite{breiman2001random} to perform 10-fold cross-validation on the 20,000 apps considered above.
Each app is represented by a feature vector where the package name of each common library is taken as a feature. 
Recall that we have collected 892 libraries from those 20,000 apps.
Therefore, our machine learning-based experiments contain 892 features.
Table~\ref{tab:ml} illustrates the results of our machine-learning-based malware detection through 4 different settings, which are

\begin{itemize}

\item \textbf{E1: } Machine-learning experiments using the entire set of 20,000 apps.

\item \textbf{E2: } Some apps do not contain any of our harvested libraries, leading to empty feature vector which can lead to mis-classifications. Thus, we conduct an experiment taking into account only apps which include at least one of our collected common libraries.

\item \textbf{E3: } We replay the experiment E2 where we ensure that there is no class imbalance: we randomly select 5,424 apps from the 8,580 antivirus-flagged apps.

\item \textbf{E4: } Similarly to experiment E3, we repeat E2 with a balanced dataset by oversampling the Benign set, using the Synthetic Minority Oversampling TEchnique (SMOTE)~\cite{chawla2002synthetic}.
\end{itemize}

The results of all experiments, showed in Table~\ref{tab:ml}, indicate good discriminative power of library features for machine-learning based detection of AV-flagged apps.

\begin{table}[h!]
    \centering
    \caption{Results of our machine-learning-based detection of AV-flagged apps.}
    \begin{tabular}{ l c c c c c }
	\hline
Exp.		&	Benign set		& AV-flagged set		& Precision	& Recall		& F-Measure \\   
\hline     
E1		& 	10,000		& 10,000			& 0.841		& 0.835		& 0.835		\\
E2		& 	5,424		& 8,580			& 0.861		& 0.861		& 0.861		\\
E3		& 	5,424		& 5,424			& 0.862		& 0.860		& 0.860		\\
E4		& 	8,580		& 8,580			& 0.875		& 0.873		& 0.873		\\
\hline
\end{tabular}
    \label{tab:ml}
\end{table}



\vspace{0.2cm}
\begin{tabular}{|p{3.1in}|}
\hline {\em 
Benign apps use significantly less common libraries than AV-flagged apps. The combinations of libraries in apps can be discriminated between benign and AV-flagged apps.
}\\ \hline
\end{tabular}

\subsection{RQ2: Improvements to Analysis Approaches}
In this section, we discuss two cases where our harvested common libraries show significant improvements to the performance of Android analysis approaches.

\subsubsection{\bf Piggybacking detection}
Recall that in Section~\ref{sec:example}, we have shown that piggybacking detection approaches 
are likely to yield false positives and false negatives if code contributed by common libraries are not taken into account. We provide more empirical evidence of such threats.

For our experiments we rely on a set of pairs of apps that we have collected and regrouped into two categories: the first category, \emph{FNData}, contains 761 pairs of apps where the smilarity score for each pair is below 50\% while the second category, \emph{FPData}, includes 1,100 pairs of apps where the similarity score of each pair of apps is over 80\%. 
Given the previously justified threshold of 80\% for deciding on piggybacking (cf. Section~\ref{sec:example}), we assume that all pairs in \emph{FPData} are piggybacked pairs while those in \emph{FNData} are not. 
We now explore again the similarity scores of the pairs when excluding from each app the common libraries (in $CL_{91}$) they may include. 


\emph{False positives elimination.}
Among the 1,100 pairs of apps in \emph{FPData}, 1,029 (93.5\%) remained similar above the 80\% threshold. 71 (6.5\%) pairs of apps now have similarity scores below 80\%, and can no longer be supposed to be piggybacked pairs. We manually verified 10 of pairs and found that they are indeed not piggybacked.

\emph{False negatives re-classification.}
Among the 761 pairs of apps in \emph{FNData}, 110 (14\%) have higher similarity scores, among which 2 pairs are now beyond the threshold of 80\% which would allows to re-classify them as piggybacked pairs. We have manually verified and confirmed that these two pairs of apps are piggybacked pairs: one pair was previously used in our motivating example section (Fig.~\ref{fig:fn}).

\subsubsection{\bf Machine learning for Malware detection}

We investigate the case of machine-learning based approaches for Android, and study the impact of ignoring or taking into account common libraries on the accuracy of prediction.  We consider a case study based on MUDFLOW~\cite{avdiienko2015mining} and its dataset. This dataset contains sensitive data leaks information for 15,096 malicious apps and 2,800 benign apps. MUDFLOW is a relevant example as the authors have originally foreseen the problem with libraries and thus attempted to exclude a small set of ad libraries in their experiments. With our large harvested dataset of common libraries, we investigate the performance gap that can be achieved by excluding more known libraries.

MUDFLOW performs machine learning to mine apps that use sensitive data abnormally.
More specifically, MUDFLOW takes each distinct type of sensitive data leak (from pre-defined \emph{source} to \emph{sink}) and performs an one-class classification to detect abnormal apps.
One-class classification is realistic in their experimental settings with their imbalanced data set (they have much more malicious apps than benign apps).

Since our goal is not to replicate MUDFLOW (along with its sophisticated library-unrelated parameters), but to evaluate the impact of excluding common libraries for machine learning, we propose to implement a slightly simplified approach for our experiments. Unlike MUDFLOW, which constructs a training set based on benign apps and then applies it to predict unknown apps, we simply perform 10-fold cross validation in our evaluation. As we are working on the same imbalanced data, we also choose one-class classification.

We have performed four types of experiments, which are detailed below:

\begin{itemize}

\item \textbf{E5: } We evaluate on all the 15,096 malicious apps. The feature set is made up of distinct sensitive data leaks.
Instead of taking into account \emph{source} and \emph{sink} methods, each data leak is represented through the \emph{source} and \emph{sink} categories (e.g., methods like \emph{Log.i()}, \emph{Log.e()} are represented as category \emph{LOG}).

\item \textbf{E6: } This experiment has similar settings as in \textbf{E5}, except that such sensitive data leaks that are contributed by the 12 ad libraries considered by MUDFLOW are excluded. 

\item \textbf{E7: } This experiment has similar settings as in \textbf{E6}. In this case however, the excluded set of libraries is the most constrained set of 1,113 libraries harvested in our work.

\item \textbf{E8: } This experiment has similar settings as in \textbf{E7}. In this case however, the excluded set of libraries is constituted by libraries selected based on a more loose definition of libraries. The excluding set contains 5,509 libraries, which may include a number of false positives.
\end{itemize}

The results of these four experiments are shown in Table~\ref{tab:ml2}.
Comparing \textbf{E7} to \textbf{E5}, the accuracy is indeed increased, which suggests
that the presence of common libraries code could confuse machine learning-based classifier.
However, the accuracy remained the same between \textbf{E6} and \textbf{E5}, suggesting that
the MUDFLOW \emph{whitelist}, which contains 12 libraries, is too small to impact the final results.
Interestingly, with our largest set of libraries, the accuracy of \textbf{E8} decreases slightly comparing to that of \textbf{E7}.
This suggests that the precision in common library identification is important: excluding non-library code will eventually decrease the overall performance for machine learning.

\begin{table}[!h]
    \caption{Results of our machine learning based experiments performed on the data set provided by MUDFLOW~\cite{avdiienko2015mining}.}
    \label{tab:ml2}
    \centering

    \begin{tabular}{ l c l r }
\hline
Seq. 	&	\#. of Features		& Excluding Libs	&	Accuracy	\\
\hline
1		&	109					& 0		&	81.85\%\\
2		&	109					& 12	 (MUDFLOW~\cite{avdiienko2015mining})	& 	81.85\%\\
3		&	109					& 1,113 ($t_p=0.9$,$t_a=0.1$)	& 	83.10\%\\
4		&	108					& 5,509 ($t_p=0.6$,$t_a=0.4$)	&	83.01\% \\
\hline
    \end{tabular}
\end{table}

\vspace{0.2cm}
\begin{tabular}{|p{3.1in}|}
\hline {\em 
It is possible to reduce false positive and false negative rates for piggybacking detections and malware prediction by excluding libraries based on a comprehensive whitelist. These case studies suggest that library code can mislead Android analysis, and our harvested set of common libraries can indeed be used to improve state-of-the-art approaches' performance.
}\\ \hline
\end{tabular}

\subsection{RQ3: Completeness of our harvested ad libraries}
VirusTotal is a free service that hosts about 40 antivirus products for analyzing suspicious files, including Android apps. Along with entirely malicious apps, VirusTotal is also able to identify adware and provide information in the labels. However, AV labels are not homogeneous, and there is no standard for naming malware and adware.
After manually inspecting several results of VirusTotal, we have observed seven keywords (cf. Table~\ref{tab:7keywords}) that are commonly leveraged by VirusTotal AV to tag adware.


\begin{table}[!h]
    \caption{The seven keywords (without case-insensitive matching) that we manually observed for inferring adware from the results of VirusTotal.}
    \label{tab:7keywords}
    \centering

    \begin{tabular}{ l l l l l l l}
    \hline
    \scriptsize
adware 	& adsware	 & addisplay & adswo  & adwo		& adrads 	& ``multi ads" \\
\hline
    \end{tabular}
\end{table}


\begin{figure}
\centering
\includegraphics[width=\linewidth]{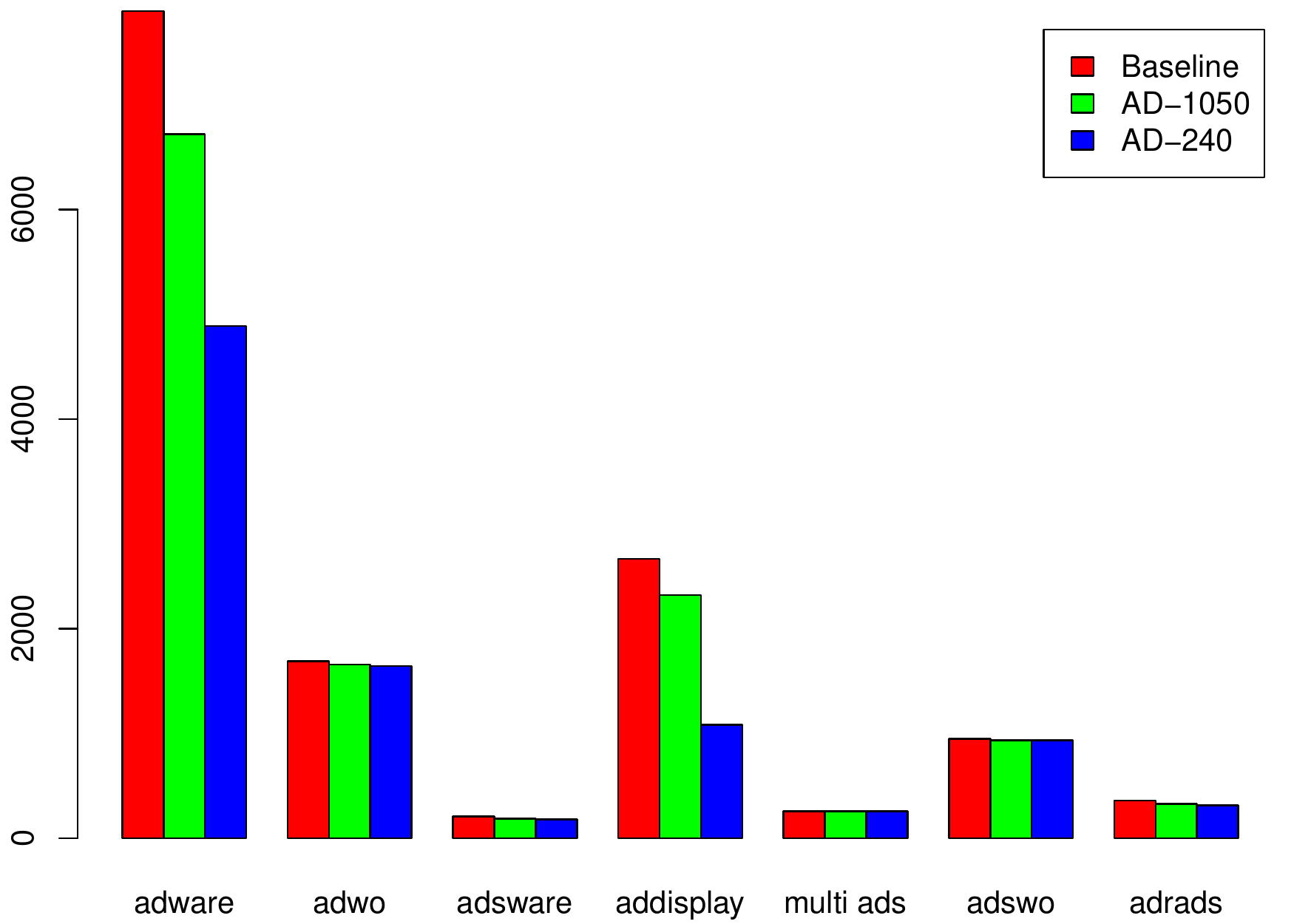}
\caption{Investigation results of comparing our ad libraries to the adware results of VirusTotal.}
\label{fig:ad2vt}
\end{figure}

In this study, we first select a set of apps that are flagged by VirusTotal as adware,
and then we inspect whether those apps could have been tagged as adware based simply on package matching with our harvested libraries. In this study, we consider 10,000 randomly sampled apps which are flagged by at least one antivirus product of VirusTotal (the flagged labels are not necessarily for adware).
Among the 10,000 apps, 8,120 (81.2\%) of them are flagged as adware following the keywords described above.
Based on the two ad sets that we have harvested before, we are able to flag 5,045 of them for $AD_{240}$ and 6,916 of them for $AD_{1050}$ as adware, giving a completeness of 62\% and 82\%, respectively.

Fig.~\ref{fig:ad2vt} presents the fine-grained results, categorized through different ad-keywords.
Our harvested ad libraries perform almost perfectly for five keywords out of the total seven keywords.
However, the performance on ``adware" and ``addisplay" keywords are less stable, indicating
that our harvested ad libraries are still missing  some less widespread libraries.

\vspace{0.2cm}
\begin{tabular}{|p{3.1in}|}
\hline {\em 
Although our harvested ad libraries are currently the largest publicly accessible set, 
it is not yet complete enough to cover all ad packages.
Nevertheless, we believe that our harvested libraries still constitute a significant set for other to  boost analysis.
}\\ \hline
\end{tabular}

\section{Threats to Validity and Discussion}
\label{sec:discussion}
Because there is no convention for specifying that a code package represents a library, 
identifying Android common libraries is challenging. We were able to perform our study 
by mining about 1,5 million apps collected over several months. Our study however 
presents a few threats to validity:

Currently, our approach is not fully aware of obfuscation, which may lead to incomplete results.
However, our findings are based on a large datasets of apps to reduce the influence of obfuscated apps.
Besides, our findings could be leveraged in settings where for example ad-libraries are represented by features which are resilient to obfuscation (e.g., called SDK API methods). In future work, we plan to conduct experiments for validating this possibility.

In this work, we do not take into account the different versions of libraries.
Thus, the validity of our similarity analysis could be threatened, as the similarity of two versions of a library could vary significantly. 
However, as our experiments are done on a large scale of apps, this threaten is somewhat mitigated by the variety and the scale of our dataset.

Our investigation into libraries have also revealed interesting findings on the use of libraries:
\begin{itemize}
	\item well-used libraries, such as \emph{unit3d}, are often used as the compromising point for malicious apps.
	\item Malware writers often name their malicious components after famous and pervasively used libraries from reputed firms: e.g., the \emph{DroidKungFu} malware family spreads malicious payload within a package called \emph{com.google.update}. Our similarity analysis allowed to detect such fraud by further investigating outliers.
\end{itemize}

\section{Related Work}
\label{sec:related_work}

At first, we discuss a batch of works that investigate the issues related to libraries.
Then, we show that even if libraries are not harmful by themselves, they threaten the validity of other approaches.
Finally, we summarize the works that are dedicated to the identification of Android libraries.

\textbf{Problems of Libraries.}
As reported by Hu et al.~\cite{hu2014duet}, Android libraries are currently suffering three threats:
1) the library modification threat, where normal libraries can be modified to be malicious.
Our previous work has also confirmed this findings~\cite{li2015ungrafting}.
2) the masquerading threat, e.g., a well-known malware family called \emph{DroidKungFu} uses names such as \emph{com.google.update} to pretend the services are provided by Google~\cite{zhou2012dissecting}.
3) the aggressive library threat, where some legitimate libraries have aggressive behaviors such as collecting users' email address.

Other works~\cite{li2014automatically, li2015iccta, book2013case} done by us and by others, have also shown that some libraries frequently and aggressively collect (leak) users' private information.
For instance, the most common leaked information is the device id, which is used by ad libraries to uniquely identify a user.
This findings are in line with the investigation of Stevens et al.~\cite{stevens2012investigating}, in which the authors show that, through libraries, users can be tracked by a network sniffer across ad providers and by an ad provider across apps.
Besides, they also argue that ad libraries usually require permissions beyond their real needs and some bad programmed libraries use Android's Javascript extension mechanism insecurely.
AdRisk~\cite{grace2012unsafe} focuses on detecting privacy and security risks posed by ad libraries.
Most notably, it shows that some libraries even execute untrusted code from internet sources.
Moreover, those untrusted code are fetched through an unsafe mechanism, which by itself has caused serious security risks.

Gui et al.~\cite{gui2015truth} have shown that free ad libraries actually come with hidden cost for developers such as the rating of apps.
As reported by Mojica et al.~\cite{mojica2014impact}, ad libraries are indeed impacting the ratings of Android apps.

Although our work in this paper is not dedicated to identify problems of libraries, our findings, the list of common (ad) libraries, can definitely benefit other approaches (e.g., API studies~\cite{linares2014mining, linares2013api, linares2014api, bavota2015impact}) by giving them a good starting point for thorough analysis.

\textbf{Research Works threatened by Libraries.}
Researchers have noticed Android libraries will definitely influence the results of app clone detection~\cite{wang2015wukong, chen2014achieving, crussell2012attack, crussell2013scalable, linares2014revisiting}, most of them use a list of libraries as a whitelist.
Detecting and filtering third-party libraries for clone detection is important, as the results may be doomed if the studied apps are dominated by common libraries.
Chen et al.~\cite{chen2014achieving} leverage a whitelist containing 73 libraries in their approach, which is far away from being a complete whitelist of existing libraries,
as shown in~\cite{wang2015wukong}, over 600 distinct libraries have been identified.
However, this list is not publicly available.
Besides, comparing to our findings in this paper, this list is also considerably incomplete.

Not only for clone detection, but also for machine learning-based malware detection, the results are threatened by common libraries.
MUDFLOW~\cite{avdiienko2015mining}, as an example, uses a list of 12 well-known ad libraries as a whitelist, to exclude such features that fill in them.
A later work done by Li et al.~\cite{li2015potential} also leverage that list in their machine learning-based malware detection.

Our work, in this paper, provides a comprehensive list of common libraries, that can be leveraged by other approaches and thus to significantly refine their results.

\textbf{Identification of Libraries.}
Wang et al.~\cite{wang2015wukong} uses an automated clustering technique to detect common libraries, in which they have found over 600 distinct libraries.
Our approach is in line with their assumptions on common libraries, however, we come with a different approach  and we also discriminate ad libraries from common libraries, for which they have not.
Another approach called AdDetect~\cite{narayanan2014addetect}, identifies Android ad libraries through their semantics (e.g., the usage of Android components, or specific APIs) and then performs a ML-based classification to detect ad libraries.
However, this approach does not report any findings that can benefit the Android research community.

\section{Conclusion}
\label{sec:conclusion}
We have presented our process for collecting a set of 1,113 common libraries and 240 ad libraries from a dataset of about 1.5 million Android apps. To the best of our knowledge, these two sets are the largest ones that are publicly accessible in the community of Android research. 

We empirically illustrate how these two library sets can be used as \emph{whitelist}s by Android analysis approaches to improve their performances. More specifically, we have shown that two approaches, namely piggybacking detection and machine learning-based malware detection, can indeed benefit from our harvested libraries.

\balance
\bibliographystyle{plain}
\bibliography{andsec}

\end{document}